\documentclass{birkart04}

 \theoremstyle{definition}

 \theoremstyle{remark}

 \numberwithin{equation}{section}

\begin{document}

%--------------------------------------------------------------------------

% editorial commands: to be inserted by the editorial office

%

%\firstpage{1}

%\volume{228}


%\DOI{003-0001}


%\seriesextraline{This is the Concrete Title of this Book\br H.E. Rowe and S.T.C. Wore, Eds.}

%


%




%\commby{inhouse GRRR}





%---------------------------------------------------------------------------


%

\title[Introduction to Step Dynamics and Step Instabilities ]{Introduction to Step Dynamics \\
and Step Instabilities}

%----------Author 1

\author[Joachim Krug]{Joachim Krug}

\address{%
Universit\"at zu K\"oln \\
Institut f\"ur Theoretische Physik \\
Z\"ulpicher Strasse 77 \\
50937 K\"oln, 
Germany}

\email{krug@thp.uni-koeln.de}

\thanks{This work was supported in part by DFG within 
SFB 237 \textit{Unordnung und grosse Fluktuationen} and 
SFB 616 \textit{Energiedissipation an Oberfl\"achen}.}




\date{\today}

\begin{abstract}

This paper provides an elementary introduction to the basic concepts
used in describing epitaxial crystal growth in terms of the thermodynamics
and kinetics of atomic steps. Selected applications to morphological
instabilities of stepped surfaces are reviewed, and some open problems
are outlined.
\end{abstract}
%%% ----------------------------------------------------------------------

\maketitle

%%% ----------------------------------------------------------------------

%\tableofcontents

\section{Introduction}

The hallmark of a crystalline solid is the discrete translational symmetry
imposed by the crystal lattice. At the surface of a crystal, this leads to the
existence of atomic \textit{steps}, which separate exposed 
lattice planes (\textit{terraces}) that differ in height
by a single lattice spacing. At sufficiently low temperatures, steps are
thermodynamically stable, in the sense that the creation of a new step segment
entails a positive free energy cost per unit length of the step. Thus steps are
long-lived surface defects, which makes them suitable as a basis for the description
of the surface morphology on an intermediate (\textit{mesoscopic}) scale, between atomistic and   
macroscopic levels of modeling.  

In crystal growth, steps play a central role because they provide the \textit{kink} sites
at which new atomic units are incorporated into the crystal (see Fig.1). The growth of
a crystal surface can thus be reduced to the advancement of existing steps, the 
nucleation of new closed step loops (i.e., atomic height \textit{islands}), and the annihilation
of steps by the merging of islands and terraces. This point of view was pioneered by 
Burton, Cabrera and Frank (BCF) in their classic 1951 paper \cite{Burton51}. The exchange
of matter between the steps and the population of adsorbed atoms (\textit{adatoms}) on the
terraces implies that the step motion has to be formulated as a moving boundary problem. 
With the advent of powerful numerical techniques for the solution of such problems,
step dynamics has become an attractive alternative to atomistic modeling approaches
in epitaxial crystal growth. Recent progress along these lines is described in many
contributions to this volume. 

The main purpose of the present article is to introduce
the basic concepts used in the step-dynamical approach to crystal growth. 
Since this approach is most useful when the number of steps is conserved,
i.e. in the absence of island nucleation and merging, we will largely restrict
ourselves to the growth on stepped (\textit{vicinal}) surfaces.  
In Sects.\ref{Thermodynamics} and \ref{Kinetics} we will present the fundamental thermodynamic
and kinetic notions in a reasonably self-contained manner, with considerable attention
to physical assumptions and approximations that enter into the construction of the model.  
Sections \ref{Stability} and \ref{Nonlinear} provide a brief overview of applications
to morphological instabilities of stepped surfaces. Here the discussion will be rather
cursory, but key references will be given to guide the reader into the extensive
literature. In the final section we try to formulate
some challenges for the future. 
General introductions to the subject can be found, e.g., in 
\cite{Pimpinelli98,Politi00,Michely03}.

\section{Thermodynamics of steps}
\label{Thermodynamics}

The fundamental abstraction of the step-dynamical approach is the replacement
of the microscopic step conformation, which is a (possibly very convoluted) path
on the discrete surface lattice, by a smooth curve in continuous space. 
At least locally, the step can then be described by the graph of a function 
$y(x)$ (Fig. 1).

\subsection{Step free energy and step stiffness}
\label{freeenergy}

The thermodynamic properties of a single step are contained in the step free energy per unit
length $\delta$ which, because of the underlying crystal structure, is generally a function 
$\delta(\theta)$ of the in-plane step orientation angle $\theta$ (see Fig.1). The step is
a long-lived, thermodynamically stable object as long as $\delta > 0$. At the thermal
roughening transition of the surface the step free energy vanishes. Above the roughening
transition temperature $T_R$ steps proliferate and a description of the surface in terms
of isolated steps and flat terraces is no longer possible. The concept of a roughening
transition was first conceived by BCF \cite{Burton51}, but the nature of the transition
was clarified only much later, in the 
seventies of the past century \cite{vanBeijeren87,Nozieres91}. For
applications to epitaxial growth the roughening transition is of little importance,
as the growth processes of interest usually proceed far below $T_R$.

\begin{figure}[htb]
\begin{center}
\includegraphics[width=\textwidth]{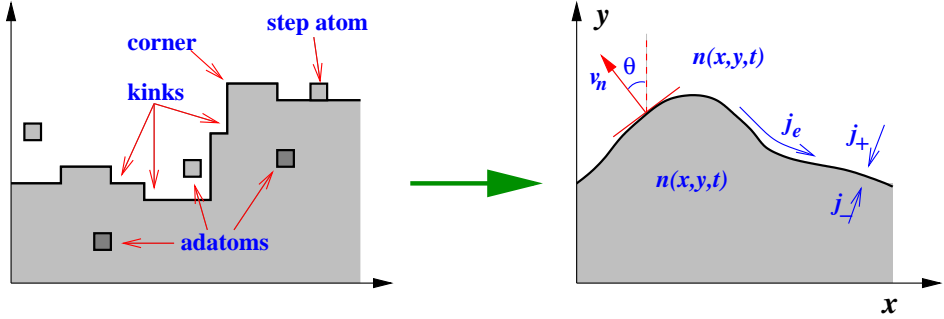}
\end{center}
\caption[]{\textit{Left panel}: 
Sketch of a step on a crystal surface of fourfold symmetry. The upper
terrace is shaded. The step consists of straight (close packed) segments separated by kinks
and corners. Adatoms diffuse on the terraces and can be incorporated into the crystal by
attaching at kinks sites. An atom moving along a straight segment of the step is called a 
step atom. \textit{Right panel}: In the step-dynamical treatment the discrete lattice step is replaced
by a smooth curve $y(x)$. The local step orientation is described by the angle $\theta$
between the step normal and the $y$-axis. The step moves at speed $v_n$ along the normal
direction. The adatom diffusion field $n(x,y,t)$ is defined on both the upper and lower
terrace, but it may be discontinuous across the step (see Fig.\ref{bcfig}). Step atoms contribute to the
step edge current $j_e$, while terrace adatoms attaching to, detaching from, or crossing
the step are contained in the mass currents $j_+$ and $j_-$.}
\label{abstract}
\end{figure}

Given the function $\delta(\theta)$, the free energy $F$
of an arbitrary step configuration $y(x)$ is obtained by integrating along the step,
\begin{equation}
\label{free}
F = \int ds \; \delta(\theta(s)) = \int dx \; \sqrt{1 + (dy/dx)^2} \; \delta(\theta(x)),
\end{equation}
where $s$ denotes the arc length along the step and $\theta(x) = 
\mathrm{arctan}(dy/dx)$. From (\ref{free}) the \textit{step chemical potential},
i.e., the free energy change upon adding an atom to the step, 
can be derived by functional differentiation. This yields the expression
\begin{equation}
\label{stepchem}
\mu = \Omega \frac{\delta F}{\delta y} = \Omega \tilde \delta \kappa,
\end{equation}
where $\Omega$ denotes the area occupied by a surface atom, $\kappa$ is the 
step curvature, and the quantity 
$\tilde \delta = \delta + d^2 \delta/d \theta^2$ 
is known as the \textit{stiffness} of the step. The step stiffness appears in a local
version of the Wulff construction, which relates the step free energy  
to the equilibrium shape of two-dimensional islands
\cite{Michely03,Nozieres91}. In this formulation the
equilibrium condition reads simply 
\begin{equation}
\label{localeq}
\tilde \delta \kappa = \textrm{const.}, 
\end{equation}
i.e. the local curvature of the equilibrium shape
is inversely proportional to the local step stiffness. 

The calculation of $\delta(\theta)$ for a specific system requires the application of equilibrium
statistical mechanics to a microscopic model of the surface. A popular microscopic model is the
Ising lattice gas, where each site of the surface lattice is either vacant or occupied and the
energy of a configuration is obtained by summing over all pairs of occupied nearest neighbor sites.
A simple, model-independent expression for the step stiffness is obtained for
steps along the close-packed (= minimal energy)
directions of the lattice at low temperatures. Such steps
consist of straight close-packed segments interspersed by a small concentration
of kinks. Elementary statistical mechanics considerations then show that \cite{Michely03}
\begin{equation}
\label{stiff}
\tilde \delta \approx \frac{k_{\textrm{B}} T}{2a} e^{\epsilon/k_{\textrm{B}} T},
\end{equation} 
where $k_{\textrm{B}}$ denotes Boltzmann's constant, $T$ is the temperature, $a$ the
in-plane lattice spacing, and $\epsilon$ is the energy cost of a kink.
     
\subsection{Step roughness}
\label{Roughness}

The existence of a roughening transition for two-dimensional surfaces, which was briefly
mentioned in the preceding
subsection, is due to the fact that steps are extended, one-dimensional objects. As a consequence
the free energy of a step increases with its length (provided $\delta > 0$), and the 
free energy cost for the spontaneous creation of a macroscopic piece of a step becomes
prohibitively large. In contrast, the creation of a kink on a one-dimensional step requires
only a fixed energy amount $\epsilon$, and hence a step contains a finite kink concentration
at any nonzero temperature. This implies that the step is \textit{rough} whenever
$T > 0$: A typical step conformation looks like the graph of a one-dimensional random walk,
with a diffusivity proportional to the kink concentration. Assuming that the step is on
average aligned with the $x$-axis, the roughness can be quantified by the height-difference
correlation function 
\begin{equation}
\label{stepdiffuse}
\langle [y(x) - y(x')]^2 \rangle = \frac{k_{\textrm{B}} T}{\tilde \delta} \vert x - x' \vert.
\end{equation}
Here the angular brackets refer to a thermal average with respect to the equilibrium
distribution. The relation (\ref{stepdiffuse}) has been widely exploited to derive,
via (\ref{stiff}), experimental estimates for the kink energy $\epsilon$ from scanning tunneling
microscopy observations of step fluctuations \cite{Jeong99,Giesen01}. 

A thermodynamic consequence of step roughness is that
the step free energy $\delta(\theta)$ cannot contain any 
singularities. Via the Wulff construction \cite{Burton51,Nozieres91}, 
this implies that the corresponding equilibrium island shapes have finite
curvature everywhere [compare to (\ref{localeq})], 
and hence display no corners or ``facets'' in the thermodynamic
limit (of course, islands of a size comparable to or smaller than the mean distance
between kinks nevertheless have atomically straight edges).  
This is in contrast to the three-dimensional case, where cusps in the 
Wulff-plot of the \textit{surface} free energy lead to facets in the equilibrium
crystal shape \cite{Nozieres91,Bonzel03}.   
 
\subsection{Step-step interactions}

\label{Interactions}

So far we have considered a single atomic step in isolation. To understand the thermodynamics
and kinetics of stepped surfaces, which consist of an array of parallel steps of equal sign
(see Fig.\ref{modes}),
it is crucial to take into account also the interactions between the steps. An important
source of interaction is the very fact that steps are thermally rough. As was first noted
by Gruber and Mullins \cite{Gruber67}, in an array of parallel steps the thermal 
wandering of a given step is impeded by collisions with its neighbors (since the steps cannot
cross). This reduces the entropy of the step, thus increasing its free energy and 
leading to an effective repulsion between the steps. 

It is straightforward to estimate
the strength of the repulsion from (\ref{stepdiffuse}). Suppose the average distance between
the steps (along the perpendicular $y$-direction)
is $l$. Then the \textit{collision length} $L_c$, the distance between close
encounters of two steps measured along the parallel $x$-direction, is obtained
by setting the left hand side of (\ref{stepdiffuse}) equal to $l^2$ and inserting
$L_c$ for $\vert x - x' \vert$. This yields $L_c \approx \tilde \delta l^2/(k_{\textrm{B}} T)$.
Assuming a fixed entropy loss $k_{\textrm{B}} C$ per collision, the free energy of the
step is then increased by an amount
\begin{equation}
\label{interaction}
\Delta \delta \approx \frac{C k_{\textrm{B}} T}{L_c} = \frac{C (k_{\textrm{B}} T)^2}{\tilde \delta 
l^2}.
\end{equation}
The coefficient $C$ can be evaluated for a model in which the steps are represented as world
lines of non-interacting fermions (subject only to the non-crossing constraint), which yields
$C = \pi^2/6$ \cite{Joos91}. 

The important feature of (\ref{interaction}) is that the step interactions decay quite slowly
with step distance, as $l^{-2}$. There is extensive experimental support for this interaction
law from scanning tunneling microscopy measurements of the distribution of step spacings on 
vicinal surfaces \cite{Jeong99,Giesen01}. 
However, the interaction strength is typically found to be considerably
larger than predicted by the purely entropic expression (\ref{interaction}). This is 
because elastic interactions between the strain fields associated with the steps, mediated
through the bulk crystal, display the same $l^{-2}$ decay law, and typically dominate
the entropic interactions. Electrostatic dipoles associated
with the steps also lead to an $l^{-2}$-interaction.
A macroscopic consequence of the $l^{-2}$-repulsion is the
singular behavior $z(y) - z(y_0) \sim (y - y_0)^{3/2}$ of the height of the equilibrium
crystal shape $z(y)$ near a facet edge at $y = y_0$ \cite{vanBeijeren87,Nozieres91,Bonzel03}.

\section{Step dynamics as a moving boundary problem}
\label{Kinetics}

Within the step dynamical picture, a step is a mathematically sharp boundary that evolves
by exchanging mass with the continuous adatom concentration field $n(x,y,t)$ on the terraces.
Thus the formulation of a step dynamical model requires, first, the specification of the
dynamics of the adatom concentration, and, second, the formulation of boundary conditions
for $n$ at the steps. Whereas the first part is quite straightforward, the second contains
a number of (explicit and implicit) assumptions about the underlying microscopic physics. 
As the formulation of the appropriate boundary conditions  
constitutes a central part of the step-dynamical modeling approach,
we will discuss these issues in considerable detail.

\subsection{Adatom dynamics}

We want to describe physical situations where atoms arrive at the surface with a deposition
flux $F$, diffuse over the terraces with a surface diffusion coefficient $D$, 
and possibly desorb back into the vacuum at rate $1/\tau$. In some cases it is also of
interest to include a directed force $\vec{f}$
acting on the adatoms due to an electric current
in the bulk of the sample (\textit{surface electromigration}) \cite{Stoyanov90,Yagi01,Minoda03}. 
Together
these processes imply the evolution equation 
\begin{equation}
\label{adatoms}
\frac{\partial n}{\partial t} = D \nabla^2 n - \frac{D}{k_{\textrm{B}} T} {\vec{f}}\cdot \nabla n 
+ F - \frac{n}{\tau}  = 
- \nabla \cdot \vec{J} + F - \frac{n}{\tau}
\end{equation}
for the adatom concentration field. Here the adatom current $\vec{J}$ has been introduced,
which is generally driven by the concentration gradient as well as by the electromigration
force. The coefficient $D/k_\textrm{B} T$ in front of the drift term is the adatom mobility, which 
is related to the diffusion coefficient through the Einstein relation.

The motion of the steps is often slow compared
to the time scale on which the adatom concentration changes, so that (\ref{adatoms}) can be
solved quasistatically, setting $\partial n/\partial t = 0$. To see when this approximation
is justified, consider for concreteness the case of growth, and 
suppose that the typical distance between steps is $l$. Then
the step velocity is of the order $v \sim F \Omega l$, which is to be compared
to the diffusion velocity $v_D \sim D/l$, the effective speed of an atom diffusing
over the distance $l$ \cite{Ghez88}. 
The quasistatic approximation holds if $v \ll v_D$ or 
\begin{equation}
\label{Peclet}
l \ll (D/\Omega F)^{1/2}. 
\end{equation}
This condition is
essentially always fulfilled, because nucleation of new islands creates steps at
typical distances $l_D \sim (D/F)^\chi$, where $\chi < 1/2$ \cite{Michely03}, and hence the step 
density is usually much larger than required according to (\ref{Peclet}).
The situation is different in the presence of an electromigration force, because then also
the drift velocity $v_\textrm{drift} = (D/k_\textrm{B} T) \vert \vec{f} \vert$ 
of the adatoms would be required to be
large compared to $V$; since the electromigration force is very weak, this condition
is not necessarily satisfied \cite{PierreLouis03} (see also Sect.\ref{lengths}). 
In the following we nevertheless adhere to the quasistatic approximation.

\subsection{Kinetic processes at the step}

\begin{figure}[hbt]
\begin{center}
\includegraphics[width=\textwidth]{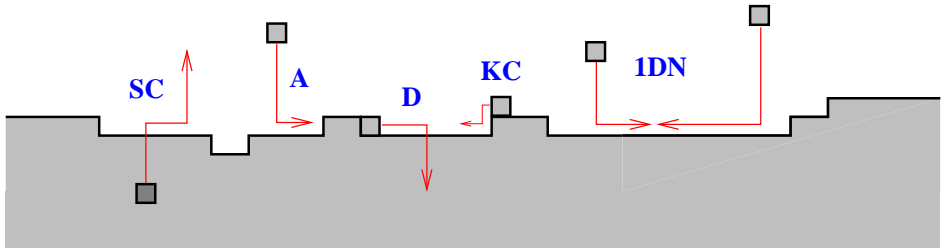}
\end{center}
\caption[]{Atomic processes at a step: Attachment (A) and detachment (D) of terrace atoms; 
step crossing (SC); kink crossing (KC); and one-dimensional nucleation (1DN).
As in Fig.\ref{abstract}, the upper terrace is shaded.}
\label{processes}
\end{figure}

The interaction of adatoms with steps involves the following microscopic processes
\cite{Roland92,Caflisch99,Filimonov04} (see Fig.\ref{processes}):

\begin{itemize}

\item \textit{Attachment and detachment.} It is important to note that these processes
actually consist of two stages: When attaching to the step, an adatom first attaches to a straight step
segment and then moves along the segment until it reaches a kink, where it is incorporated;
similarly a detachment event requires first that an atom detaches from a kink to the 
straight step, and
subsequently detaches from the step segment onto the terrace.

\item \textit{Step crossing.} The foregoing remark implies that an atom may also
cross a step without attaching to a kink; this happens if the kink concentration is low
and the binding to the straight step weak, so that the step atom detaches from the straight
step before a kink is encountered. 

\item \textit{One-dimensional nucleation} \cite{Voronkov70}. If two step atoms are present simultaneously on
a straight segment of the step, they can meet and form a step dimer, which is essentially
a pair of kinks of opposite sign. This provides a nonequilibrium mechanism for the creation of kinks,
in addition to the thermally excited kinks that are present in equilibrium (see Sect.
\ref{Roughness}). If step atoms cannot detach onto the terrace, the typical distance between
kinks created by one-dimensional nucleation is
\begin{equation}
\label{l1d}
l_{\textrm{1d}} \sim \left( \frac{D_\textrm{e}}{F_{\textrm{1d}}} \right)^{1/4},
\end{equation}
where $D_\textrm{e}$ is the diffusion coefficient of a step atom along a straight step segment,
and $F_{\textrm{1d}}$ is the one-dimensional flux impinging onto the step from the terrace
\cite{Bartelt92,Villain92}. Generalizations of (\ref{l1d}) to other conditions can be 
found in \cite{Roland92,Voronkov70,Kallunki02}.

\item \textit{Diffusion along a rough step.} Since steps always contain a finite concentration
of kinks (of equilibrium or nonequilibrium origin), mass transport by step edge diffusion requires
that step atoms are able to cross kinks and corners. The diffusion along a rough step is therefore
considerably slower than the diffusion along a straight step segment. As indicated
in Fig.\ref{processes}, a step atom crossing a kink ``from above'' first has to round
the kink and then detach from it onto the step. Kink rounding is often associated
with an additional energy barrier \cite{PierreLouis99,Murty99,Kallunki03}.  

\end{itemize}

\subsection{Linear constitutive relations}

Within the step dynamical model, the diverse and rather complex processes described in
the preceding subsection are lumped
together into the net mass currents $j_+$ and $j_-$ from the lower and upper terrace,
respectively, and the edge diffusion current $j_e$ (see Fig.\ref{abstract}). 
In the spirit
of linear nonequilibrium thermodynamics, these currents are assumed to be linear
in the adatom concentration gradients and differences. Consider first the edge diffusion current
$j_e$, which counts the number of atoms that pass a point on the step in unit time.
It is of the form
\begin{equation}
\label{je}
j_e = -\sigma \frac{\partial \mu}{\partial s},
\end{equation}
a relation that can be taken to define the \textit{mobility} $\sigma$ of an edge atom
along a (rough, curved) step \cite{Kallunki03,Krug95}. It is important to keep in mind that 
(\ref{je}) includes only that part of the mass transport along the step that is induced by 
differences in the step chemical potential, and hence serves to restore thermal equilibrium at the
step. In addition, there are genuinely nonequlibrium contributions to the current that arise from the
combined effects of attachment, step edge diffusion 
and kink rounding barriers \cite{PierreLouis99,Murty99,Politi00a,Rusanen02}. 

\begin{figure}[hbt]
\begin{center}
\includegraphics[width=0.7\textwidth]{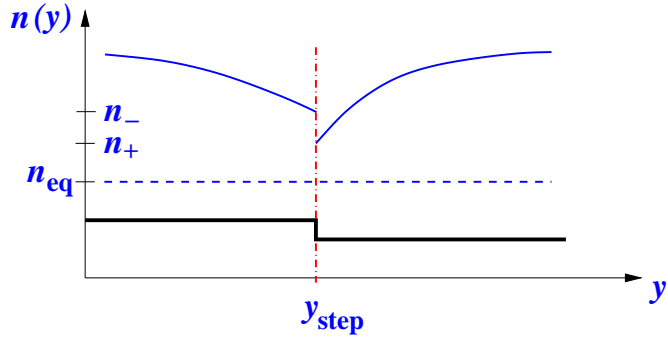}
\end{center}
\caption[]{Behavior of the adatom concentration near a step.}
\label{bcfig}
\end{figure}

Figure \ref{bcfig} illustrates how the adatom concentration changes across a step.  
At each point of the step, three different values of $n$ can be defined:
The limiting value $n_+$ ($n_-$) attained by the terrace concentration field $n(x,y)$ when approaching
the step from the lower (upper) terrace, and the equilibrium step atom concentration $n_{\textrm{eq}}$
which would be established in the absence of growth, through the detachment of atoms
from thermally excited kinks. The mass currents $j_+$ and $j_-$ that leave a terrace via the ascending
and descending step, respectively, are now written as linear combinations of the differences between
these adatom concentrations, in the form (see e.g. \cite{PierreLouis03})
\begin{equation}
\label{genbc+}
j_+ = k_+(n_+-n_{\textrm{eq}}) + p (n_+-n_-)
\end{equation}
and
\begin{equation}
\label{genbc-}
j_- = k_-(n_--n_{\textrm{eq}}) - p (n_+-n_-).
\end{equation}
The first term on the right hand side of (\ref{genbc+}) is the net current of terrace atoms that
attach to the step from below, while the second term is the current of atoms that cross the
step without attaching to it. The first contribution is driven by the deviation of the adatom
concentration from its equilibrium value, while the second contribution is driven by the
difference of the adatom concentrations on the two terraces. Similar considerations apply to
the current (\ref{genbc-}) from the upper terrace. The kinetic coefficients $k_+$ and
$k_-$ are known as \textit{attachment rates}, while the coefficient $p$ is referred to as
the \textit{step permeability}; see Sects.\ref{Schwoebel} and
\ref{Permeability} for further discussion of physical effects related
to these quantities. Usually all kinetic coefficients are assumed to be positive,
but negative values can also be meaningful under certain conditions \cite{Zhao04}
(see Sect.\ref{Electromigration}). The 
symmetry between the step crossing terms in (\ref{genbc+}) and (\ref{genbc-}) can be viewed
as a consequence of the Onsager reciprocity relations \cite{PierreLouis03}.   

The linearity of the constitutive relations (\ref{je},\ref{genbc+},\ref{genbc-}) contains the implicit
assumption that the step is close to equilibrium. There have been recent attempts to 
formulate kinetic models that do not require this assumption \cite{Caflisch99,Filimonov04},
but the bulk of the work in the field relies on it. Detailed kinetic modeling is also needed
to derive expressions for $k_\pm$ and $p$ in terms of 
the rates of elementary atomic processes\footnote{For the step edge mobility $\sigma$ in 
(\ref{je}) such a derivation has been achieved, see \cite{Kallunki03,Krug95}.}.
Here we follow the common practice and regard the kinetic coefficients in (\ref{genbc+},\ref{genbc-})
as phenomenological parameters.

\subsection{Thermodynamic driving force}

The thermodynamic driving forces that are
responsible for restoring equilibrium enter (\ref{genbc+},\ref{genbc-})
through the equilibrium adatom concentration $n_{\textrm{eq}}$, which is related to the (local)
step chemical potential through the standard thermodynamic identity 
\begin{equation}
\label{neq}
n_\textrm{eq} = n_\textrm{eq}^0 e^{\mu/k_\textrm{B} T} \approx   n_\textrm{eq}^0(1 + 
\mu/k_\textrm{B} T).
\end{equation}
Here $n_\textrm{eq}^0$ is the equilibrium adatom concentration at an isolated, straight step, and the
second relation assumes that the deviations from this value are small. As was
discussed in Sect.\ref{Thermodynamics}, the step chemical potential is affected by the
step curvature [see (\ref{stepchem})] and by the step-step interactions. To derive
the latter contribution, we denote by $y_j(x)$ the position of the $j$-th step in the
$(x,y)$-plane, and assume (following Sect.\ref{Interactions}) that the 
free energy of the 
step-step interaction (including both entropic and direct, elastic or electrostatic contributions) 
for two steps at a distance $l$ is of the form $V(l) = A/l^2$, where $A$ is the interaction
strength. Then the contribution to the step chemical potential of step $j$ from the interaction
with the neigboring steps $j-1$ and $j+1$ is 
\begin{equation}  
\label{muint}
\Delta \mu_j^{\textrm{int}} = \frac{\partial}{\partial y_j} [ V(y_{j+1} - y_j) + V(y_j - y_{j-1})]
\end{equation}
and together with the curvature contribution the expression for the chemical potential of step
$j$, to be inserted in (\ref{neq}), is
\begin{equation}
\label{mutot}
\mu_j = \Omega \tilde \delta \kappa_j + 2A \left[ \frac{1}{(y_{j+1} - y_{j})^3} - \frac{1}{(y_{j} - 
y_{j-1})^3} \right],
\end{equation}
with $\kappa_j$ denoting the curvature of step $j$. Note that all terms in (\ref{mutot}) also
depend on the coordinate $x$ along the step.  

\subsection{Mass conservation}

Once the mass currents $j_\pm$ have been fixed through the constitutive relations 
(\ref{genbc+},\ref{genbc-}), the boundary conditions for the adatom concentration field
as well as the evolution equations for the step follow simply from mass conservation.
This requires, first, that the mass currents are continuous at the step, which means
that the total terrace current $\vec{J}$ appearing on the right hand side of (\ref{adatoms})
must be matched to the local currents $j_\pm$ at the steps. Thus
\begin{equation}
\label{curr+}
j_+ = D \vec{n} \cdot \left[\nabla n \vert_+ - \frac{1}{k_\textrm{B} T} \vec{f} n \right]
\end{equation}
and   
\begin{equation}
\label{curr-}
j_- = - D \vec{n} \cdot \left[\nabla n \vert_- - \frac{1}{k_\textrm{B} T} \vec{f} n \right],
\end{equation}
where $\nabla n \vert_\pm$ are the values of the concentration gradient on the two sides
of the step, and $\vec{n}$ is the unit vector normal to the step, which points towards
the lower terrace. Together with (\ref{genbc+},\ref{genbc-}), these equations constitute
the boundary conditions for the solution of the (quasistatic) drift-diffusion equation
for $n$.

The second consequence of mass balance is that the step moves in response to the net
attachment current $j_+ + j_-$ (note that the contributions due to step crossing cancel),
as well as due to the mass transport along the step. The normal velocity of the step is
therefore given by
\begin{equation}
\label{vn}
v_n = \Omega \left( j_+ + j_- -\frac{\partial j_e}{\partial s} \right).
\end{equation}
This completes the derivation of the step-dynamical model.

\subsection{The Ehrlich-Schwoebel effect}
\label{Schwoebel}

The attachment rates $k_\pm$ were introduced by Schwoebel in his seminal 1969 paper 
\cite{Schwoebel69}. Schwoebel and Shipsey were the first to explore the consequences
of attachment asymmetry ($k_+ \neq k_-$) for the stability of growing stepped surfaces
\cite{Schwoebel69,Schwoebel66}. 
Their work was triggered by the field ion microsope observation of 
Ehrlich and Hudda that tungsten adatoms on tungsten surfaces
tend to be reflected by descending step edges \cite{Ehrlich66},
which implies that $k_+ > k_-$. Therefore the preferential attachment
to ascending steps, which is observed for
many materials, is 
often referred to as the \textit{(normal) Ehrlich-Schwoebel (ES) effect}. 

For metal surfaces, the ubiquity of the normal ES effect is now fairly well established
\cite{Michely03}, but for semiconductor materials the situation is less clear. For example,
for the Si(111)-surface, a detailed microscopy study of island nucleation near steps 
\cite{Chung02} provides
evidence for an \textit{inverse} ES effect ($k_+ < k_-$) at high temperatures, above
the transition from the $(7 \times 7)$ to the $(1 \times 1)$ reconstruction, 
but at lower temperatures (below the transition) a study of electromigration-induced step
and island motion \cite{Saul02} indicates that, on the contrary, $k_+ > k_-$. 

\subsection{Step permeability}
\label{Permeability}

The importance of step crossing processes for the evolution of the surface morphology has been
realized only fairly recently. Step permeability in the sense of (\ref{genbc+},\ref{genbc-}) was
first introduced in 1992 by Ozdemir and Zangwill \cite{Ozdemir92}. The dimensionless ratio
$p/k_+$, which is a measure for the probability that an atom crosses the step without
beging incorporated, can be quite large for semiconductor surfaces, where the incorporation process
may be difficult because of reconstructions at the step. From an analysis of the relaxation of 
biperiodic gratings on Si(001) it was concluded that $p/k_\pm = 40 \pm 20$ 
for this system \cite{Tanaka97}.
But also steps on metal surfaces can be highly permeable, as evidenced by the dramatic
formation of steep pyramids on Al(110) \cite{Mongeot03}, 
which cannot be explained without substantial
transport across steps.
In the fully permeable limit, $p \to \infty$, the finiteness of the current in 
(\ref{genbc+},\ref{genbc-}) forces the adatom concentration to be \textit{continuous} at
the steps. Some consequences of step permeability will be described in Sect.\ref{Electromigration}.

\subsection{Kinetic lengths}
\label{lengths}

Inserting the expressions (\ref{genbc+}) and (\ref{genbc-}) on the right hand sides of (\ref{curr+})
and (\ref{curr-}), the boundary conditions are seen to identify gradients of the adatom concentration $n$,
multiplied by $D$, with differences in $n$, multiplied by kinetic coefficients. As a consequence,
the ratios of $D$ to the kinetic coefficients naturally define the length scales 
\begin{equation}
\label{kinlengths}
l_+ = \frac{D}{k_+}, \;\;\; l_- = \frac{D}{k_-}, \;\;\; l_0 = \frac{D}{p},
\end{equation}
which are collectively referred to as \textit{kinetic lengths}. The kinetic lengths are useful
in discussing the effects of step boundary conditions on the morphological evolution. For
example, in assessing the importance of a conventional ES effect in growth, it is crucial to compare
the typical terrace sizes to the kinetic length $l_-$; the ES effect is \textit{strong},
and makes itself felt during the growth of the first few layers, if $l_-$ is large compared
to the terrace size \cite{Politi00,Michely03} (see also Sect.\ref{Straight}). 

Similarly the comparison of the two terms inside the square brackets in (\ref{curr+}) and (\ref{curr-})
leads to the definition of the \textit{electromigration length} \cite{PierreLouis00}  
\begin{equation}
\label{xi}
\xi = \frac{k_\textrm{B} T}{\vert \vec f \vert},
\end{equation}
and the discussion of non-quasistatic effects can be phrased in terms of a 
``P\'eclet-length'' 
\begin{equation}
\label{xiPe}
\xi_P = \frac{D}{v}, 
\end{equation}
where $v$ is the step velocity \cite{PierreLouis03}.
This is motivated by the fact that going to a frame moving with speed $v$ introduces
a convection term $v \nabla n$ in the stationary diffusion equation, which is of the
same form as the electromigration term in (\ref{adatoms}). In general, the quasistatic
approximation requires that all other relevant length scales are small compared
to $\xi_P$.  

\section{Morphological stability}
\label{Stability}

The solution of the moving boundary problem formulated in the preceding section
in its full generality is a formidable challenge. For this reason much of the work
in this area has been restricted to linear stability analyses of simple surface 
morphologies. The most important results will be summarized in this section.

\subsection{Stability of growing and sublimating step trains}

\begin{figure}[hbt]
\begin{center}
\includegraphics[width=0.9\textwidth]{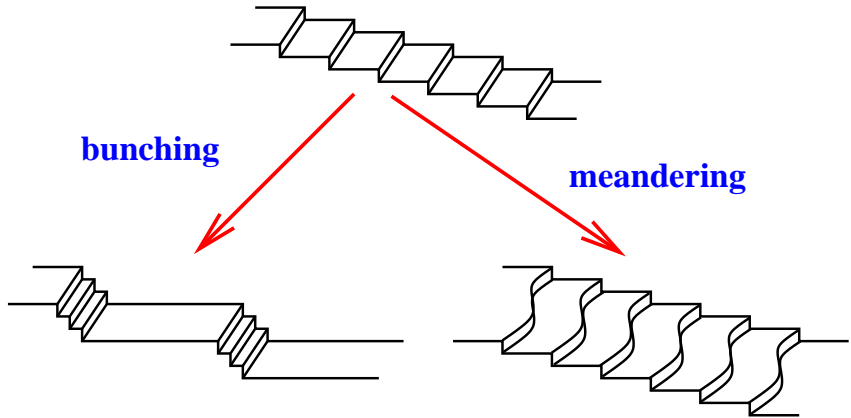}
\end{center}
\caption[]{Instability modes of a vicinal surface.}
\label{modes}
\end{figure}

The basic instability modes of a regular step train -- a vicinal surface with
straight, equidistant steps -- are illustrated in Fig.\ref{modes}: Either the
surface separates into regions of high step density (\textit{step bunches}) and 
wide terraces, or the steps become wavy (\textit{step meandering}). While the 
two modes are usually assumed to be mutually exclusive, there is experimental 
evidence from homoepitaxial growth that they may also coexist \cite{Neel03,Neel03b}. 

\subsubsection{Bunching of straight steps.}
\label{Straight}

The stability analysis is particularly simple if the steps can be assumed to remain
straight, because then the dynamical problem reduces to a set of ordinary differential equations
coupled along the $y$-direction. The basic stability results for straight, impermeable steps were
derived by Schwoebel and Shipsey \cite{Schwoebel69,Schwoebel66}, who showed that
a growing step train is stabilized by a normal ES effect ($k_+ > k_-$) while
the same effect leads to step bunching during sublimation; conversely, an inverse
ES effect ($k_+ < k_-$) implies step bunching during growth. 
Explicitly, the equations of motion for the step positions $y_j$ during growth are
of the form
\begin{equation}
\label{eqsmotion}
\frac{d y_j}{dt} = f_+(y_{j+1} - y_j) + f_-(y_j - y_{j-1}),
\end{equation}
where $f_+$ and $f_-$ denote the contributions from the leading and the trailing terrace,
respectively, which are given by 
\begin{equation}
\label{fpm}
f_{\pm}(l) = \frac{F \Omega l}{2} \frac{2 \l_{\mp} + l}{l_++l_-+l}.
\end{equation}
Here the effects of step-step interactions arising from the 
step chemical potential (\ref{mutot}) have been neglected.
A straightforward linear stability analysis of (\ref{eqsmotion}) shows that
the equidistant step train is stable provided $d [f_+(l)-f_-(l)]/dl > 0$, which,
using the explicit expression (\ref{fpm}), is seen to be equivalent to
$k_+>k_-$. The functional form of (\ref{fpm}) illustrates the role 
of the kinetic lengths $l_\pm$ in gauging the strength
of the attachment asymmetry: When the terrace size $l \gg l_\pm$ the attachment
kinetics becomes effectively symmetric and $f_+ \approx f_- \approx F \Omega l/2$.     

\begin{figure}[hbt]
\begin{center}
\includegraphics[width=0.9\textwidth]{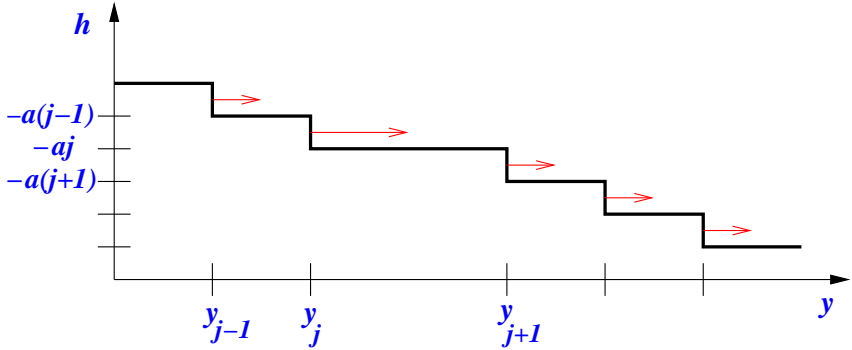}
\end{center}
\caption[]{Stabilization of a configuration of equidistant, straight steps
by a normal ES effect. 
Step $j$ receives most of its flux
from the leading terrace $y_j < y < y_{j+1}$. It therefore
moves faster than the neighboring steps, and the 
step spacing is equalized. In our notation the terrace in front
of step $j$, measured from an arbitrary reference level, 
is at height $h = -aj$, where $a$ is the (vertical) lattice constant.
}
\label{1dsteps}
\end{figure}

The stabilization 
of the growing step train by the normal ES effect
is illustrated in Fig.\ref{1dsteps}.
Formally it can be interpreted in terms of an effective step-step repulsion
mediated by the diffusion field on the terraces. This repulsion is very efficient, in the
sense that it leads to terrace width fluctuations that are much smaller than the
corresponding fluctuations in thermal equilibrium \cite{Krug95b,PierreLouis98a}.

The fact that a normal ES effect stabilizes the regular step spacing implies
that step bunching during growth -- which is actually observed rather frequently in 
experiments -- requires a separate mechanism for its explanation. It has been realized
for a long time that impurities may cause step bunching during growth \cite{Cabrera58,Kandel94,Krug02}.
Recently it was pointed out by Pimpinelli and coworkers that step bunching can also be induced
if the diffusing species that is incorporated during growth (e.g., the adatoms) is coupled
to a second species, which could be a chemical precursor in the case of chemical vapor
deposition \cite{Pimpinelli00} or highly mobile dimers \cite{Vladimirova01}. Using a two-species
generalization of BCF theory, it can be shown that under suitable conditions a
normal ES effect for the second species implies an effective inverse ES effect
for the growth species, and hence causes step bunching. On reconstructed surfaces such as 
Si(001) the pronounced anisotropy of terrace diffusion can also cause step bunching
\cite{Myslivecek02}. Finally, through a subtle coupling to a preceding meandering instability,
fast edge diffusion has been predicted to provide a mechanism for step bunching \cite{Politi00a}.
This last scenario seems consistent with the combined meandering and bunching instability
observed for Cu(100) \cite{Neel03,Neel03b}.

\subsubsection{Step meandering.}

Bales and Zangwill (BZ) first showed that a normal ES effect induces step meandering during
growth \cite{Bales90}. Because of the effective step-step repulsion, the meander can be most
easily accomodated if the steps meander in phase, as indicated in Fig.\ref{modes}. Within the
linear stability analysis, this implies that the in-phase meander is the mode with the largest
growth rate \cite{Pimpinelli94}. 
A fundamental result of the BZ analysis is a prediction for the meander
wavelength in the initial stage of the instability. In the absence of desorption
but taking into account edge diffusion, the relevant expression reads \cite{Gillet00}
\begin{equation}
\label{lambdaBZ}
\lambda_{\textrm{BZ}} = 4 \pi \sqrt{\frac{(D n_{\textrm{eq}}^0 l/k_{\textrm{B}} T + \sigma)
\Omega \tilde \delta}{F l^2 f_{\textrm{ES}}}}.
\end{equation}
Here $l$ is the step spacing of the vicinal surface, which is assumed to be
small compared to the meander wavelength, and the factor 
$f_\textrm{ES} = (l_- - l_+)/(l_++l_-+l)>0$
is a measure for the strength of the ES effect. Equation (\ref{lambdaBZ}) makes it clear
that the meander wavelength is determined by the competition between the destabilizing flux
in the denominator, and the stabilizing thermodynamic forces represented by the step stiffness
in the numerator; shifting the balance towards the stabilizing/destabilizing side increases/decreases
the meander wavelength. The stiffness is multiplied by the sum of two kinetic coefficients representing
the two kinetic pathways that contribute to smoothening the step, terrace diffusion and step
edge diffusion.

\begin{figure}[hbt]
\begin{center}
\includegraphics[width=0.8\textwidth]{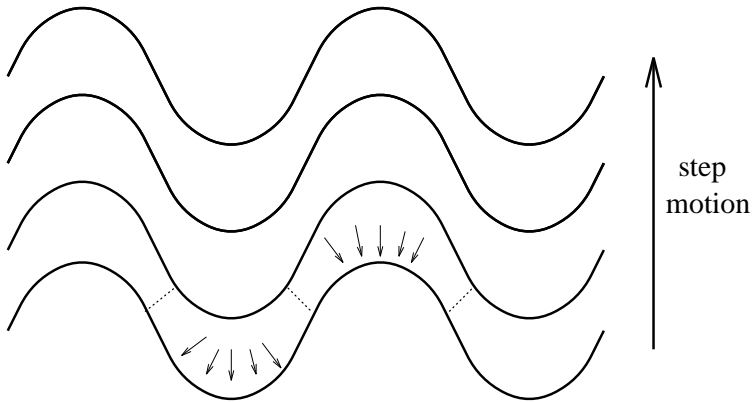}
\end{center}
\caption[]{Geometric origin of the Bales-Zangwill meandering instability
\cite{Kallunki03a,Krug02}. The terraces are subdivided into lots
along the dotted lines, which are drawn perpendicular to the lines of constant adatom concentration.
Each lot receives the same number of atoms per unit time, which, owing to the ES effect,
attach primarily to the corresponding segment of the ascending step. Because of the meander,
the indented segments of the step are longer than the protruding ones. Since both receive
the same total flux, the protrusions propagate faster and the meander is amplified.}
\label{BZ}
\end{figure}

The basic mechanism underlying the BZ instability is 
well known from more conventional diffusional interface instabilities, e.g. in solidification
\cite{Caroli91}: The preferential attachment of adatoms to the ascending step implies
that the growth of protrusions in the step is amplified. In contrast to solidification, however,
here we are dealing with an infinite array of coupled ``interfaces''.
The situation is illustrated  in Fig.\ref{BZ}. It is clear from this figure that the essential
feature needed for the instability is that the flux reaching the step from the lower terrace
is larger than the corresponding flux from the upper terrace. Such an imbalance can be 
achieved even without an attachment asymmetry, by simply making the lower terrace larger than
the upper terrace. Experiments on suitably tailored step structures on the Si(111) surface \cite{Homma01}
(for which the existence of a ES effect is controversial, see Sect.\ref{Schwoebel})
do indeed produce meander patterns, and confirm the prediction of the BZ linear stability
analysis that the meander wavelength should scale with the inverse square root of the flux
[see (\ref{lambdaBZ})]. By the same argument, step meandering can result if the
upper and lower terraces differ, e.g., in the diffusion or desorption rates for adatoms,
a situation that is relevant to growth on Si(111) near the $7 \times 7 \to 1 \times 1$
transition temperature \cite{Kato04}.
   
A detailed experimental study
of the temperature and flux dependence of the meander wavelength
was reported recently for growing surfaces vicinal to Cu(100) 
\cite{Neel03b,Maroutian01,Maroutian01a}. The main conclusion of this work was that the experimentally
observed meander instability \textit{cannot} be attributed to the BZ mechanism; instead,
it seems that quantitative agreement can be reached \cite{Kallunki02,Rusanen01} if an alternative
mechanism associated with kink rounding barriers (the 
\textit{Kink Ehrlich Schwoebel Effect} or KESE) \cite{PierreLouis99,Murty99,Politi00a} 
is assumed to be operative. 
The KESE is the one-dimensional analog of the ES effect \cite{Neel03b}.
In the present case it is triggered by the disturbances created
by one-dimensional nucleation events on the perfectly straight steps. Consequently
the initial meander wavelength is given by the nucleation length (\ref{l1d}). This changes
the exponent of the flux dependence as compared to (\ref{lambdaBZ}), and leads to 
a temperature dependence governed by the activation energy for step edge diffusion.

In contrast to the BZ instability, the KESE acts independently
at each step, and hence there is no specific phase relation between the deformations of
different steps in the initial regime of the instability \cite{Kallunki02}. Phase coherence
develops later due to the effective step-step repulsion induced by the ES effect or,
in the absence of any attachment asymmetry, due to entropic or energetic step interactions;
particularly in the last case, the synchronization of the step meanders is found in simulations
to be very slow \cite{Kallunki04a}. 

Step meandering can also occur in the absence of either a ES effect or a KESE,
if edge diffusion is sufficiently fast. The phenomenon was discovered in kinetic Monte Carlo simulations
\cite{Murty99}, and it can be understood within the framework of a coarse grained
continuum (``height'') description of the surface, based on the notion of 
growth-induced surface currents \cite{Politi00,Michely03,Krug02,Krug04,Rost04}
(see also Sect.\ref{Nonlinear}).
The starting point of the 
height description is a suitable phenomenological nonlinear partial differential
equation for the surface height field $h(x,y,t)$.   
The stability analysis for a growing vicinal surface carried out in this seetting
shows that the surface is unstable towards step meandering whenever there
is a net surface current in the uphill direction \cite{Krug95,Rost96}.  
Such a current can be argued to arise by step edge diffusion for purely geometric
reasons, if the steps are 
already somewhat wavy; then the motion of an atom along a step, directed on average from
protrusions to indentations, also has an uphill component.
Because it presupposes a certain step corrugation, the effect is, in a sense, nonlinear, and
it has so far not been possible to treat it within the conventional stability analysis of 
the step dynamical model. Kinetic Monte Carlo simulations indicate that the meander
wavelength scales with flux as $\lambda \sim F^{-1/4}$, but a theoretical understanding
of this result is lacking \cite{Videcoq02}. 

\subsection{Stability of islands}

We have emphasized in the preceding subsection that the BZ instability requires a
larger flux to the step from the lower, compared to the upper, terrace. It is clear
that this situation also arises during the growth of two-dimensional islands on a 
\textit{singular} surface (a surface without preexisting steps). When the islands
nucleate their radius is much smaller than the distance between islands \cite{Pimpinelli98,Michely03},
and hence the adatoms that contribute to the growth of the island originate mainly from
the substrate, i.e. from the lower terrace relative to the island boundary. Conversely, 
in the late stage of growth close
to island coalescence, the island boundary is mainly fed from above, as the uncovered areas
between islands are smaller than the islands themselves. One therefore expects that the island
boundary should be unstable to meandering during the early stages of growth, and that it is
restabilized when coalescence is approached. Of course it is also possible that the island
boundary remains stable, if the island size is small compared to the characteristic
meander wavelength [given e.g. by (\ref{lambdaBZ})] throughout the growth history. 

This qualitative picture is confirmed by linear stability analyses. 
The first study of island stability was carried out by Avignon and Chakraverty, who considered
a single island growing in the presence of desorption \cite{Avignon69}; in this case the
length scale for the island size at which restabilization sets in is given by the
diffusion length 
\begin{equation}
\label{xs}
x_s = \sqrt{D \tau}, 
\end{equation}
which is a measure for the range of correlations
induced by the diffusion field. The first mode that becomes unstable
during growth is a deformation of threefold symmetry\footnote{Crystal anisotropy
of the stiffness or the kinetic coefficients is not taken into account
in \cite{Avignon69,Bales95}.}. A self-consistent treatment of island stability in the presence
of surrounding islands was performed by Bales and Chrzan \cite{Bales95}, who arrived
at similar conclusions.
Initial instability followed by restabilization has been observed experimentally for
micron-sized islands on Si(111) \cite{Homma01}.

\subsection{Stability under surface electromigration}
\label{Electromigration}

In 1989, Latyshev and collaborators reported that a direct heating current can induce step bunching
on surfaces vicinal to Si(111) \cite{Latyshev89}. The bunches dissolve upon reversing the current
direction, which suggested the hypothesis \cite{Stoyanov90} that a current-driven 
directed motion of silicon
adatoms -- \textit{surface electromigration} -- may be responsible for the phenomenon\footnote{For a 
general discussion of electromigration effects see \cite{Kuhn04} and references therein.}. 
In a seminal paper, Stoyanov
included the electromigration force into the BCF theory and showed that step bunching indeed
occurs, if the force is in the down-step direction \cite{Stoyanov91}. Subsequent experimental
work has confirmed the electromigration hypothesis, but at the same time a bewildering 
diversity of electromigration-induced patterns on vicinal silicon surfaces has been discovered
\cite{Yagi01,Minoda03}. 

In particular, the current direction needed to induce step bunching was found to reverse three (!)
times with increasing temperature. The scenario first proposed by Stoyanov \cite{Stoyanov91} appears    
to apply to the lowest temperature regime (\textit{regime I}), and much theoretical work has been devoted to 
trying to understand the reversals at higher temperatures. The simple idea that the reversals
may be attributed to a sign change of the effective charge of the adatoms \cite{Kandel96}
has been ruled out by experiments \cite{Degawa00} which show that the direction of the force coincides
with the current direction at all temperatures. Other mechanisms that have been proposed to 
explain the reversals include strong desorption, in the sense that the diffusion length
(\ref{xs}) becomes comparable to the step spacing \cite{Stoyanov91}, and diffusion and drift of
surface vacancies \cite{Misbah95}.

While the issue is far from being settled, recent work seems to converge on the view that
the resolution of the puzzle must be sought in the boundary conditions for the adatom
concentration, at least as far as the first reversal (from regime I to regime II) is concerned. 
Stoyanov first showed that step bunching occurs, in the presence of sublimation, 
for a force in the up-step direction,
if the steps are assumed to be perfectly permeable ($l_0 = 0$ in (\ref{kinlengths})) 
\cite{Stoyanov97,Stoyanov98}. Moreover, he predicted that the  
unstable current direction should reverse in the presence of a growth flux, in 
agreement with subsequent experiments \cite{Metois99}; at least in the
absence of an attachment asymmetry, no such dependence on growth or
sublimation conditions is present for impermeable steps \cite{PierreLouis03}.  
Further support for the permeability picture comes from the observation of 
in-phase step meandering in 
regime II, when the current is in the down-step direction \cite{Minoda01}, a phenomenon that is
reproduced by the linear stability analysis for permeable steps \cite{Sato00}. 

Nevertheless
alternative explanations cannot be ruled out. For example, it has been argued that the
consequences of fast attachment kinetics, in the sense of $l_+, l_- \ll l$, can hardly
be distinguished from those of step permeability \cite{PierreLouis03}. 
Evidence for step bunching due to an uphill force in the case of fast attachment
kinetics has been found in kinetic Monte Carlo simulations, and the effect has
been reproduced in a linear stability analysis assuming an electric field dependence of the
kinetic coefficients and the equilibrium adatom concentration in the boundary
conditions (\ref{genbc+},\ref{genbc-}) \cite{Suga00}. A particularly
elegant recently proposed scenario for the stability reversal 
is based on the observation that the effective 
kinetic coefficients $k_\pm$ entering
the boundary conditions may become \textit{negative}, if the step is treated as a
spatially extended region in which the diffusion of adatoms is enhanced rather
than hindered relative to diffusion on the terraces \cite{Zhao04}. Indeed, noting that the basic
stability criterion derived by Stoyanov in his 1991 paper \cite{Stoyanov91} involves the
\textit{product} of the electromigration force and the attachment coefficient,
it is clear that a sign reversal of the kinetic coefficient is tantamount to
a reversal of the direction of the force. 

Additional complications, not discussed here, arise when the current direction is varied
continuously relative to the orientation of the steps \cite{Krug04,Dobbs96,Degawa01}.  

\section{Nonlinear evolution}
\label{Nonlinear}

\subsection{Step bunching}

Numerical integration of the step dynamical equations in the nonlinear regime has been
largely restricted to the bunching of straight steps, where the problem reduces
to a one-dimensional array of coupled ordinary differential equations;
representative examples out of a large number of publications are \cite{Stoyanov98b,Liu98,Sato99,Sato01}.
The main interest in these studies has been to characterize the shape of individual
bunches, as well as the time evolution of the typical bunch size due to bunch
coarsening and coalescence. For both aspects quantitative information is available
from experiments on electromigration-induced step bunching of surfaces
vicinal to Si(111) \cite{Yang96,Fujita99,Homma00}. 

Some of the bunch properties can be 
derived analytically by passing from the discrete step dynamics
to a continuum description of the one-dimensional height profile $h(y,t)$
perpendicular to the steps \cite{Krug04}. 
The continuum limit is rather straightforward
for the case of step flow growth, 
because the creation and annihilation of steps by island nucleation and coalescence
does not have to be taken into account. We sketch the derivation
for a particularly simple situation, thus providing an illustration
of how the step dynamical model can be related to the continuum
height equations discussed elsewhere in this volume \cite{Rost04}.

Specifically, consider step bunching in the regime of a strong inverse
ES effect, such that the typical step spacing $l$ satisfies
$l_+ \gg l \gg l_-$. Then the equations of motion (\ref{eqsmotion},\ref{fpm})
become \textit{linear},
\begin{equation}
\label{linear}
\frac{dy_j}{dt} = \hat F (y_j - y_{j-1}),
\end{equation}
where $\hat F = F \Omega$. Clearly
$y_j(t)$ can be replaced by a continuous function $y(h,t)$,
where $h = - a j$ (compare to Fig.\ref{1dsteps}), by simply expanding
the difference on the right hand side of (\ref{linear}). To second
order this yields\footnote{The expansion has to be carried to third
order in the symmetric case $k_+ = k_-$ \cite{Krug97}.}
\begin{equation}
\label{linearcont}
\frac{\partial y}{\partial t} = \hat F \left[ -a \frac{\partial y}{\partial h} - 
\frac{a^2}{2} \frac{\partial^2 y}{\partial h^2} \right].
\end{equation}
Applying the chain rule in the form 
\begin{equation}
\label{chain}
\frac{\partial h}{\partial t} = - \frac{\partial h}{\partial y} \frac{\partial y}{\partial t}
\end{equation}
appropriate for the negatively sloped surface, we obtain a nonlinear evolution 
equation for $h(y,t)$ \cite{Krug97},
\begin{equation}
\label{hevol}
\frac{\partial h}{\partial t} + \frac{\partial}{\partial y} \left[ - \frac{\hat Fa^2}{2} 
\left( \frac{\partial h}{\partial y} \right)^{-1} \right] = \hat F a.
\end{equation}
The right hand side of this equation is naturally interpreted as a conservation
law with a \textit{growth-induced current} \cite{Politi00,Michely03,Krug02,Krug04,Rost04}, 
which is inversely proportional
to the surface slope and directed downhill (recall that $\partial h/\partial y < 0$).

The step-step interaction terms arising from (\ref{mutot}) can be handled in a similar
way \cite{Krug04b}. Putting together the destabilizing and
stabilizing effects, the resulting surface evolution equation takes the general form
\begin{equation}
\label{hevolfull}
\frac{\partial h}{\partial t} + \frac{\partial}{\partial y} \left[ - A 
\left( \frac{\partial h}{\partial y} \right)^{-1} + B 
\left( \frac{\partial h}{\partial y} \right)^{-1} \frac{\partial^2}{\partial y^2} 
\left( \frac{\partial h}{\partial y} \right)^2 
 \right] = \mathrm{const.}
\end{equation}
where the coefficients $A$ and $B$ are positive, and the constant on the right hand side
is the deposition flux. The same evolution equation has been obtained for step bunching induced by 
a normal ES effect during sublimation \cite{Krug04} and for electromigration-induced
step bunching in the limit of slow attachment/detachment kinetics
($l_\pm \gg l$) \cite{Liu98}. It applies whenever the destabilizing
part of the step dynamical equations, which is responsible for the step bunching
instability, becomes linear in the step spacings [as in (\ref{linear})]. 

The scaling properties of stationary step bunches can be derived from (\ref{hevolfull}) by setting
the current inside the square brackets to a constant and investigating the resulting ordinary
differential equation for the slope $\partial h/\partial y$ \cite{Liu98,Krug04b,Nozieres87}. One
finds, in particular, that the minimal terrace size $l_{\mathrm{min}}$ inside the step bunch
(the inverse of the maximal slope) scales with the number of steps $N$ in the bunch as
$l_{\mathrm{min}} \sim N^{-2/3}$, which is close to the behavior observed in experiments
\cite{Fujita99,Homma00}. At present it is not clear to what extent continuum equations
like (\ref{hevolfull}) are capable of also describing the time-dependent behavior of step 
bunches \cite{Krug04}. 

Equation (\ref{hevolfull}) is a representative of a larger class of evolution equations which
have been proposed within a general classification scheme for step bunching phenomena
\cite{Pimpinelli02}. All these equations are highly nonlinear, in the sense that the
typical nonlinearities have the form of inverse powers of the slope. A different kind of 
evolution equations with polynomial nonlinearities arises when the derivation is carried
out close to the threshold of the step bunching instability \cite{Gillet01}.

Finally, we note that all continuum equations discussed so far in this subsection have
the form of a local conservation law [broken only by a constant flux on the right hand
side, see (\ref{hevol}) and (\ref{hevolfull})]. One might expect that this
reflects the absence of desorption from the surface, which generally violates 
the property of volume (hence height) conservation \cite{Krug02,Krug04,Rost04,Smilauer99}.
However, the explicit derivation of an equation of the form (\ref{hevolfull})
for sublimation-induced step bunching \cite{Krug04b} shows that the structure of the
conservation law is retained even in this case, provided desorption is a weak effect,
in the sense of the diffusion length (\ref{xs}) being large compared to the other
relevant length scales. In the presence of \textit{strong} desorption 
one obtains instead an evolution equation with polynomial singularities
of Kuramoto-Sivashinsky type, which displays spatiotemporal chaos
rather than coarsening behavior \cite{Sato95} (see \cite{Krug04} for 
a general discussion).
     
\subsection{Step meandering}
\label{NonlinearMeander}
The nonlinear evolution of growth-induced step meandering has been studied
mostly within the framework of \textit{local} approximations to the 
nonlocal moving boundary problem, which can be systematically derived using
multiscale techniques. The result of such a calculation is a one-dimensional,
nonlinear partial differential equation which (depending on the situation of 
interest) describes either a single step interacting with the diffusion field
\cite{Bena93} or the collective step coordinate of a step train with an
in-phase meander \cite{Gillet00,PierreLouis98b,PierreLouis98c}. 
As in the preceding subsection, evolution equations 
of nonconserved type \cite{Bena93,PierreLouis98b} (displaying spatiotemporal chaos) and conserved
type \cite{Gillet00,PierreLouis98c} may be distinguished. To give an example of the latter kind,
the evolution equation \cite{Kallunki00}
\begin{equation}
\label{meanderevol}
\frac{\partial y}{\partial t} = - \frac{\partial}{\partial x} \left \{
\frac{\alpha y_x}{1 + y_x^2} + \frac{\beta}{(1 + y_x^2)^\nu} 
\frac{\partial}{\partial x} \left[ \frac{y_{xx}}{(1 + y_x^2)^{3/2}}
\right] \right \}
\end{equation}
with positive coefficients $\alpha$, $\beta$
describes the collective step meander for the Bales-Zangwill instability;
the exponent $\nu$ depends on 
whether the dominant step smoothing mechanism is terrace diffusion
($\nu=1$) or step edge diffusion ($\nu=1/2$). In both cases the
equation displays unbounded amplitude growth, with $\langle y \rangle
\sim t^{1/2}$, while the lateral wavelength of the pattern remains
fixed. Related equations with other nonlinear terms may also display
unbounded lateral coarsening \cite{Politi04} or interrupted coarsening
limited to a finite time interval \cite{Danker03} (see also \cite{Krug04}).

\section{Open problems}

At the end of these introductory notes, it seems appropriate to formulate
some open questions that could be, and should be, addressed in the near
future.

\subsection{Microscopic basis of kinetic coefficients}

For the step dynamical model to attain predictive power, it is mandatory
to achieve a good understanding of the relationship between the kinetic
coefficients entering the boundary conditions and the underlying atomistic
rates. There has been encouraging progress in this direction
(see e.g. \cite{Kallunki02}), but the
problem is clearly not solved in full generality. 
Presumably it will be necessary to go 
beyond the assumption of equilibrium at the steps, along the lines of 
\cite{Caflisch99,Filimonov04}.  

\subsection{Crystal anisotropy}

In the same vein, it is essential to take account of the fact that
both thermodynamic and kinetic properties of steps on crystal
surfaces are generally strongly anisotropic; still crystal
anisotropy is ignored in most of the work in the field.
As was recently demonstrated for the case of step meandering, 
anisotropy may \textit{qualitatively} alter the evolution of 
morphological instabilities \cite{Danker03}. An attractive application
for a numerical step-dynamical 
scheme incorporating crystal anisotropy would be 
to reproduce the sequence of island shapes that is observed
experimentally as a function of temperature on metal surfaces
such as Pt(111) \cite{Michely03}.

\subsection{Exotic step instabilities}

We have briefly mentioned two novel step instabilities related to edge
diffusion, which cannot be captured
by straightforward linear stability analysis, and which thus may be termed
\textit{exotic}: The meandering instability caused by fast edge diffusion without
attachment asymmetry or kink rounding barriers \cite{Murty99}, and the step bunching
instability induced by edge diffusion on steps with a preexisting meander
\cite{Politi00a}. So far evidence for the existence of these instabilities
comes mainly from kinetic Monte Carlo simulations \cite{Murty99,Videcoq02}.
A fully nonlinear treatment within the step dynamical model (including, in particular,
the appropriate edge diffusion currents) would provide important guidance towards
an analytic understanding, which is ultimately needed to assess the relevance
of these effects for real surfaces.     

\subsection{Synchronization of the step meander}

As was discussed above in Sect.\ref{NonlinearMeander}, 
the nonlinear evolution of the step meandering instability has so far
been treated under the assumption that the in-phase meander is fully coherent,
so that the whole pattern can be described by a single step conformation
$y(x,t)$. With the availability of powerful numerical algorithm for the
solution of the full moving boundary problem for multiple steps, it will
soon be possible to address also the initial phase of the instability,
where the phase relation between different steps is established.
Gaining a better understanding of the synchronization process is important,
because the phase defects in the meander pattern have been shown to 
act as seeds for secondary instabilities which eventually lead to the 
proliferation of steps of a sign opposite to the initial vicinality,
and hence to the breakdown of step flow \cite{Rost96,Kallunki04}. Kinetic
Monte Carlo simulations indicate that the breakdown is mediated by the
formation of closed step loops deep in the fjords of the meander \cite{Kallunki04}.
To handle such topology changes within the step dynamical model will 
require the use of approaches that avoid explicit tracking of the
steps, such as phase field or level set methods, which are described
elsewhere in this volume.

% ------------------------------------------------------------------------

\subsection*{Acknowledgment}

I would like to thank Jouni Kallunki and Stoyan Stoyanov 
for fruitful interactions, and John Weeks for
making \cite{Zhao04} available prior to publication.

% ------------------------------------------------------------------------


\begin{thebibliography}{1}

\bibitem{Burton51} W.K. Burton, N. Cabrera, F.C. Frank, \textit{The Growth of Crystals
and the Equilibrium Structure of their Surfaces.} Phil. Trans. Roy. Soc. A  
\textbf{243} (1951), 299--358 .

\bibitem{Pimpinelli98} A. Pimpinelli, J. Villain, \textit{Physics
of Crystal Growth.} Cambridge University Press, 1998

\bibitem{Politi00} P. Politi, G. Grenet, A. Marty, A. Ponchet,
J. Villain, \textit{Instabilities in Crystal Growth by Atomic or Molecular
Beams.} Phys. Rep. \textbf{324} (2000), 271--404.

\bibitem{Michely03} T. Michely, J. Krug, \textit{Islands, Mounds and Atoms. Patterns
and Processes in Crystal Growth Far from Equilibrium.} Springer, Heidelberg 2004.

\bibitem{vanBeijeren87} H. van Beijeren, I. Nolden, \textit{The Roughening Transition.}
In W. Schommers, P. von Blanckenhagen (Eds.),
\textit{Structure and Dynamics of Surfaces II.} (Springer, Heidelberg 1987), pp.259--300.

\bibitem{Nozieres91} P. Nozi\`eres, \textit{Shape and Growth of Crystals.} In
C. Godr\`eche (Ed.), \textit{Solids Far from Equilibrium} (Cambridge University
Press, 1991), pp.1--154.

\bibitem{Jeong99} H.-C. Jeong, E.D. Williams, \textit{Steps on Surfaces: Experiment
and Theory.} Surf. Sci. Rep. \textbf{34} (1999), 171--294.

\bibitem{Giesen01} M. Giesen, \textit{Step and island dynamics at solid/vacuum and solid/liquid interfaces.} 
Prog. Surf. Sci. \textbf{68} (2001), 1--153.

\bibitem{Bonzel03} H.P. Bonzel, \textit{3D Equilibrium Crystal Shapes in the New Light
of STM and AFM.} Phys. Rep. \textbf{385} (2003), 1--67.

\bibitem{Gruber67} E.E. Gruber, W.W. Mullins, \textit{On the Theory of Anisotropy
of Crystalline Surface Tension.} J. Phys. Chem. Solids \textbf{28} (1967), 875--887.

\bibitem{Joos91} B. Jo\'os, T. L. Einstein, N. C. Bartelt, 
\textit{Distribution of terrace widths on a vicinal surface within the one-dimensional free-fermion model.}
 Phys. Rev. B \textbf{43} (1991), 8153--8162.

\bibitem{Stoyanov90} S. Stoyanov, \textit{Heating Current Induced Conversion between 
$2 \times 1$ and $1 \times 2$ Domains at Vicinal (001) Si Surfaces - Can it be Explained
by Electromigration of Si Adatoms?} Jap. J. Appl. Phys. \textbf{29}
(1990), L659--L662.

\bibitem{Yagi01} K. Yagi, H. Minoda, M. Degawa, \textit{Step bunching, step wandering and
faceting: self-organization at Si surfaces.} Surf. Sci. Repts. \textbf{43} (2001) 45--126.

\bibitem{Minoda03} H. Minoda, \textit{Direct current heating effects on Si(111) vicinal surfaces.}
J. Phys.-Cond. Matter \textbf{15} (2003), S3255--S3280.  
                                                                                      
\bibitem{Ghez88} R. Ghez, S.S. Iyer,\textit{The Kinetics of Fast Steps on Crystal Surfaces
and its Application to the Molecular Beam Epitaxy of Silicon.} IBM J. Res. Dev.
\textbf{32}, 804 (1988)

\bibitem{PierreLouis03} O. Pierre-Louis, \textit{Step bunching with general kinetics:
stability analysis and macroscopic models.} Surf. Sci. \textbf{529} (2003), 114--134.

\bibitem{Roland92} C. Roland, G.H. Gilmer, \textit{Epitaxy on surfaces vicinal to Si(001).
II. Growth properties of Si(001) steps}. Phys. Rev. B \textbf{46} (1992) 13437--13451.

\bibitem{Caflisch99} R.E. Caflisch, W. E, M.F. Gyure, B. Merriman, C. Ratsch,
\textit{Kinetic model for a step edge in epitaxial growth.} Phys. Rev. B \textbf{59}
(1999), 6879--6887.

\bibitem{Filimonov04} S.N. Filimonov, Yu.Yu. Hervieu, \textit{Terrace-edge-kink model
of atomic processes at the permeable steps.} Surf. Sci. \textbf{553} (2004), 133--144.

\bibitem{Voronkov70} V.V. Voronkov, \textit{The movement of an elementary step by means of 
the formation of one-dimensional nuclei.} Sov. Phys. Crystallogr. \textbf{15} (1970), 8--13.

\bibitem{Bartelt92} M.C. Bartelt, J.W. Evans, \textit{Scaling analysis of diffusion-mediated
island growth in surface adsorption processes.} Phys. Rev. B \textbf{46} (1992), 12675--12687.

\bibitem{Villain92} J. Villain, A. Pimpinelli, L. Tang, D. Wolf, \textit{Terrace sizes
in molecular beam epitaxy.} J. Phys. I France \textbf{2} (1992), 2107--2121.

\bibitem{Kallunki02} J. Kallunki, J. Krug, \textit{Competing mechanisms for step meandering
in unstable growth.} Phys. Rev. B \textbf{65} (2002), 205411.

\bibitem{PierreLouis99} O. Pierre-Louis, M.R. D'Orsogna, T.L. Einstein, 
\textit{Edge Diffusion during Growth: The Kink Ehrlich-Schwoebel Effect and Resulting Instabilities.}
Phys. Rev. Lett. \textbf{82} (1999), 3661--3664.

\bibitem{Murty99} M.V. Ramana Murty, B.H. Cooper, \textit{Instability in Molecular Beam Epitaxy
due to Fast Edge Diffusion and Corner Diffusion Barriers.} Phys. Rev. Lett \textbf{83} (1999),
352--355.

\bibitem{Kallunki03} J. Kallunki, J. Krug, \textit{Effect of kink-rounding barriers on step edge
fluctuations.} Surf. Sci. \textbf{523} (2003) L53--L58.

\bibitem{Krug95} J. Krug, H.T. Dobbs, S. Majaniemi, \textit{Adatom mobility for the solid-on-solid
model.} Z. Phys. B \textbf{97} (1995), 281--291.

\bibitem{Politi00a} P. Politi, J. Krug, \textit{Crystal symmetry, step-edge diffusion, and unstable
growth.} Surf. Sci. \textbf{446} (2000), 89--97.

\bibitem{Rusanen02} M. Rusanen, I.T. Koponen, T. Ala-Nissila, C. Ghosh, T.S. Rahman,
\textit{Morphology of ledge patterns during step flow growth of metal surfaces vicinal to fcc(001).}
Phys. Rev. B \textbf{65} (2002), 041404.

\bibitem{Zhao04} T. Zhao, J.D. Weeks, D. Kandel, \textit{A unified treatment of current-induced
instabilities on Si surfaces.} (preprint)

\bibitem{Schwoebel69} R.L. Schwoebel, \textit{Step Motion on Crystal Surfaces. II.}
J. Appl. Phys. \textbf{40} (1969), 614--618.

\bibitem{Schwoebel66} R.L. Schwoebel, E.J. Shipsey, \textit{Step Motion on Crystal Surfaces.}
J. Appl. Phys. \textbf{37} (1966), 3682--3686.

\bibitem{Ehrlich66} G. Ehrlich, F.G. Hudda, \textit{Atomic View of Surface Self-Diffusion:
Tungsten on Tungsten.} J. Chem. Phys. \textbf{44} (1966), 1039--1055.

\bibitem{Chung02} W.F. Chung, M.S. Altman, \textit{Kinetic length, step permeability,
and kinetic coefficient asymmetry on the Si(111) ($7 \times 7$) surface.} Phys. Rev. B
\textbf{66} (2002), 075338.

\bibitem{Saul02}  A. Sa\'ul, J.-J. M\'etois, A. Ranguis, \textit{Experimental evidence for an Ehrlich-Schwoebel 
effect on Si(111).} Phys. Rev. B \textbf{65} (2002) 075409. 

\bibitem{Ozdemir92} M. Ozdemir, A. Zangwill, \textit{Morphological equilibration of a facetted crystal}.
Phys. Rev. B \textbf{45} (1992), 3718--3729.
 
\bibitem{Tanaka97} S. Tanaka, N.C. Bartelt, C.C. Umbach, R.M. Tromp, J.M. Blakely,
\textit{Step Permeability and the Relaxation of Biperiodic Gratings on Si(001)}.
Phys. Rev. Lett. \textbf{78} (1997) 3342--3345.

\bibitem{Mongeot03} F. Buatier de Mongeot, W. Zhu, A. Molle, R. Buzio, C. Boragno, U. Valbusa, 
E. G. Wang, Z. Zhang,
\textit{Nanocrystal Formation and Faceting Instability in Al(110) Homoepitaxy: True Upward Adatom Diffusion 
at Step Edges and Island Corners.} Phys. Rev. Lett. \textbf{91} (2003), 016102.  

\bibitem{PierreLouis00} O. Pierre-Louis, T.L. Einstein, \textit{Electromigration of 
single-layer clusters.} Phys. Rev. B \textbf{62} (2000), 13697--13706.

\bibitem{Neel03} N. N\'eel, T. Maroutian, L. Douillard, H.-J. Ernst, \textit{From Meandering to 
Faceting, Is Step Flow Growth Ever Stable?} Phys. Rev. Lett. \textbf{91} (2003) 226103.

\bibitem{Neel03b} N. N\'eel, T. Maroutian, L. Douillard, H.-J. Ernst, \textit{Spontaneous
structural pattern formation at the nanometre scale in kinetically restricted homoepitaxy
on vicinal surfaces.} J. Phys.: Condens. Matter \textbf{15} (2003), S3227--S3240.

\bibitem{Krug95b} J. Krug, M. Schimschak, \textit{Metastability of Step Flow Growth in 1+1
Dimensions}. J. Phys. I France \textbf{5} (1995), 1065--1086.

\bibitem{PierreLouis98a} O. Pierre-Louis, C. Misbah, \textit{Dynamics and fluctuations 
during MBE on vicinal surfaces. I. Formalism and results of linear theory.} 
Phys. Rev. B \textbf{58} (1998), 2259--2275.

\bibitem{Cabrera58} N. Cabrera, D.A. Vermilyea, \textit{The Growth of Crystals from Solution.}
In: \textit{Growth
and Perfection of Crystals}, ed. by R. Doremus, B. Roberts, D.
Turnbull (Wiley, New York 1958) pp.~393--408.

\bibitem{Kandel94} D. Kandel, J. Weeks: \textit{Theory of impurity-induced step bunching.} 
Phys. Rev. B \textbf{49} (1994), 5554--5564.

\bibitem{Krug02a} J. Krug, \textit{New mechanism for impurity-induced step bunching.} Europhys.
Lett. \textbf{60} (2002), 788--794.

\bibitem{Pimpinelli00} A. Pimpinelli, A. Videcoq, \textit{Novel mechanism for the onset of morphological
instabilities during chemical vapour epitaxial growth.} Surf. Sci. \textbf{445} (2000), L21--L28.

\bibitem{Vladimirova01} M. Vladimirova, A. De Vita, A. Pimpinelli, \textit{Dimer diffusion as a 
driving mechanism of the step bunching instability during homoepitaxial growth.} 
Phys. Rev. B \textbf{64} (2001), 245420.

\bibitem{Myslivecek02} J. Myslive\v{c}ek,  C. Schelling,
F. Sch\"{a}ffler, G. Springholz, P. \v{S}milauer, J. Krug, 
B. Voigtl\"{a}nder, \textit{On the microscopic origin of the kinetic step bunching
instability on vicinal Si(001).}
Surf. Sci. \textbf{520} (2002), 193--206.

\bibitem{Bales90} G.S. Bales, A. Zangwill, \textit{Morphological instability of a terrace edge
during step-flow growth.} Phys. Rev. B \textbf{41} (1990), 5500--5508.

\bibitem{Pimpinelli94} A. Pimpinelli, I. Elkinani, A. Karma, C. Misbah, J. Villain,
\textit{Step motions on high-temperature vicinal surfaces.} J. Phys.: Condens. Matter
\textbf{6} (1994), 2661--2680.

\bibitem{Caroli91} B. Caroli, C. Caroli, B. Roulet, \textit{Instabilities of planar
solidification fronts.} In
C. Godr\`eche (Ed.), \textit{Solids Far from Equilibrium} (Cambridge University
Press, 1991), pp.155--296. 

\bibitem{Gillet00} F. Gillet, O. Pierre-Louis, C. Misbah, \textit{Non-linear evolution of the
step meander during growth of a vicinal surface with no desorption.} Eur. Phys. J. B \textbf{18}
(2000), 519--534.

\bibitem{Kallunki03a} J. Kallunki, \textit{Growth instabilities of vicinal crystal surfaces
during Molecular Beam Epitaxy.} (PhD dissertation, University of Duisburg-Essen, 2003).

\bibitem{Krug02} J. Krug, \textit{Four lectures on the physics of crystal growth.} Physica A 
\textbf{313} (2002), 47--82.

\bibitem{Homma01} Y. Homma, P. Finnie, M. Uwaha, \textit{Morphological instability of atomic
steps observed on Si(111) surfaces.} Surf. Sci. \textbf{492} (2001), 125--136.

\bibitem{Kato04} R. Kato, M. Uwaha, Y. Saito, \textit{Step wandering due to the structural
difference of the upper and lower terraces.} Surf. Sci. \textbf{550} (2004), 149--165.

\bibitem{Maroutian01} T. Maroutian, L. Douillard, H.-J. Ernst, \textit{Morphological instability
of Cu vicinal surfaces during step-flow growth.} 
Phys. Rev. B \textbf{64} (2001), 165401.

\bibitem{Maroutian01a} T. Maroutian, \textit{\'Etude exp\'erimentale d'instabilit\'es de
croissance des faces vicinales} (PhD dissertation, Universit\'e Paris 7, 2001).

\bibitem{Rusanen01} M. Rusanen, I. T. Koponen, J. Heinonen, T.
Ala-Nissila, \textit{Instability and wavelength selection during step flow growth
of metal surfaces vicinal to fcc(001).} Phys. Rev. Lett. \textbf{86} (2001), 5317--5320.

\bibitem{Kallunki04a} J. Kallunki (unpublished).

\bibitem{Krug04} J. Krug, \textit{Kinetic Pattern Formation at Solid Surfaces.} In G. Radons,
P. H\"aussler, W. Just (Eds.), \textit{Collective Dynamics of Nonlinear and Disordered
Systems} (Springer, Berlin 2004). 

\bibitem{Rost04} M. Rost, \textit{Continuum models for surface growth.} (this volume)

\bibitem{Rost96} M. Rost, P. \v{S}milauer, J. Krug,
\textit{Unstable epitaxy on vicinal surfaces.} 
Surf. Sci. \textbf{369} (1996), 393--402.

\bibitem{Videcoq02} A. Videcoq, \textit{Auto-organisation de surfaces cristallines pendant la
croissance \'epitaxiale: une \'etude th\'eorique} (PhD dissertation, Universit\'e Blaise
Pascal, Clermont-Ferrand 2002).

\bibitem{Avignon69} M. Avignon, B.K. Chakraverty, \textit{Morphological stability of a 
two-dimensional nucleus.} Proc. Roy. Soc. A \textbf{310} (1969), 277--296.

\bibitem{Bales95} G.S. Bales, D.C. Chrzan, \textit{Transition from Compact to Fractal
Islands during Submonolayer Epitaxial Growth.} Phys. Rev. Lett.
\textbf{74} (1995), 4879--4882.

\bibitem{Latyshev89} A.V. Latyshev, A.L. Aseev, A.B. Krasilnikov, S.I. Stenin,
\textit{Transformations on clean Si(111) stepped surface during sublimation.}
Surf. Sci. \textbf{213} (1989) 157--169.

\bibitem{Kuhn04} P. Kuhn, J. Krug, \textit{Islands in the Stream: Electromigration-Driven
Shape Evolution with Crystal Anisotropy.} (this volume)

\bibitem{Stoyanov91} S. Stoyanov, \textit{Electromigration Induced Step Bunching on Si Surfaces - 
How Does It Depend on the Temperature and Heating Current Direction?} Jap. J. Appl. Phys. \textbf{30}
(1991), 1--6.

\bibitem{Kandel96} D. Kandel, E. Kaxiras, \textit{Microscopic Theory of Electromigration
on Semiconductor Surfaces.} Phys. Rev. Lett. \textbf{76} (1996), 1114--1117. 

\bibitem{Degawa00} M. Degawa, H. Minoda, Y. Tanishiro, K. Yagi, 
\textit{Direct-current-induced drift direction of silicon adatoms on Si(111)-($1 \times 1$) surfaces.}
Surf. Sci. \textbf{461} (2000), L528--L536. 

\bibitem{Misbah95} C. Misbah, O. Pierre-Louis, A. Pimpinelli, \textit{Advacancy-induced step
bunching on vicinal surfaces.} Phys. Rev. B \textbf{51} (1995), 17283--17286.

\bibitem{Stoyanov97} S. Stoyanov, \textit{Current-induced step bunching at vicinal surfaces
during crystal sublimation.} Surf. Sci. \textbf{370} (1997), 345--354.

\bibitem{Stoyanov98} S. Stoyanov, \textit{New type of step bunching instability at vicinal
surfaces in crystal evaporation affected by electromigration.} Surf. Sci. \textbf{416} (1998), 
200--213.

\bibitem{Metois99} J.J. M\'etois, S. Stoyanov, \textit{Impact of growth on the stability-instability
transition at Si(111) during step bunching induced by electromigration.} Surf. Sci. \textbf{440}
(1999), 407--419.

\bibitem{Minoda01} H. Minoda, I. Morishima, M. Degawa, Y. Tanishiro, K. Yagi,
\textit{Time evolution of DC heating-induced in-phase step wandering on Si(111)
vicinal surfaces.} Surf. Sci. \textbf{493} (2001), 487--493.

\bibitem{Sato00} M. Sato, M. Uwaha, Y. Saito, \textit{Instabilities of steps induced by the drift
of adatoms and effect of the step permeability.} Phys. Rev. B \textbf{62} (2000), 8452--8472.

\bibitem{Suga00} N. Suga, J. Kimpara, N.-J. Wu, H. Yasunaga, A. Natori:
\textit{Novel Transition Mechanism of Surface Electromigration Induced Step Structure
on Vicinal Si(111) Surfaces.} Jpn. J. Appl. Phys. \textbf{39} (2000), 4412--4416.

\bibitem{Dobbs96} H. Dobbs, J. Krug, \textit{Current Induced Faceting in Theory and Simulation.}
J. Phys. I France \textbf{6} (1996), 413--430.

\bibitem{Degawa01} M. Degawa, H. Minoda, Y. Tanishiro, K. Yagi, \textit{In-phase step wandering
on Si(111) vicinal surfaces: Effect of direct current heating tilted from the step-down 
direction.} Phys. Rev. B \textbf{63} (2001), 045309.

\bibitem{Stoyanov98b} S. Stoyanov, V. Tonchev, \textit{Properties and dynamic interaction
of step density waves at a crystal surface during electromigration affected sublimation.}
Phys. Rev. B \textbf{58} (1998), 1590--1600.

\bibitem{Liu98} D.-J. Liu, J.D. Weeks, \textit{Quantitative theory of current-induced step bunching
on Si(111).} Phys. Rev. B \textbf{57} (1998), 14891--14900.

\bibitem{Sato99} M. Sato, M. Uwaha, \textit{Growth of step bunches formed by the drift
of adatoms.} Surf. Sci. \textbf{442} (1999), 318--328.

\bibitem{Sato01} M. Sato, M. Uwaha, \textit{Growth law of step bunches induced by the
Ehrlich-Schwoebel effect in growth.} Surf. Sci. \textbf{493} (2001), 494--498.

\bibitem{Yang96} Y.-N. Yang, E.S. Fu, E.D. Williams, \textit{An STM study of current-induced
step bunching on Si(111).} Surf. Sci.
\textbf{356} (1996), 101--111.

\bibitem{Fujita99} K. Fujita, M. Ichikawa, S.S. Stoyanov, \textit{Size-scaling exponents of current-induced
step bunching on silicon surfaces.} Phys. Rev. B \textbf{60} (1999), 16006--16012.

\bibitem{Homma00} Y. Homma, N. Aizawa, \textit{Electric-current-induced step bunching
on Si(111).} Phys. Rev. B \textbf{62} (2000), 8323--8329.

\bibitem{Krug97} J. Krug, \textit{Continuum Equations for
Step Flow Growth}. In D. Kim, H. Park,
B. Kahng (Eds.), \textit{Dynamics of Fluctuating
Interfaces and Related Phenomena} (World Scientific, Singapore 1997), pp.~95--113.

\bibitem{Krug04b} J. Krug, V. Tonchev, S. Stoyanov, A. Pimpinelli,
\textit{Scaling properties of step bunches induced by Ehrlich-Schwoebel barriers
during sublimation.} (in preparation)

\bibitem{Nozieres87} P. Nozi\`eres, \textit{On the motion of steps on a vicinal surface.}
J. Physique \textbf{48} (1987), 1605--1608.

\bibitem{Pimpinelli02} A. Pimpinelli, V. Tonchev, A. Videcoq, M. Vladimirova,
\textit{Scaling and Universality of Self-Organized Patterns on Unstable Vicinal Surfaces.}
Phys. Rev. Lett. \textbf{88} (2002), 206103.

\bibitem{Gillet01} F. Gillet, Z. Csahok, C. Misbah, \textit{Continuum nonlinear surface
evolution equation for conserved step-bunching dynamics.} Phys. Rev. B \textbf{63}
(2001), 241401.

\bibitem{Smilauer99} P. \v{S}milauer, M. Rost, J. Krug, \textit{Fast coarsening in 
unstable expitaxy with desorption.} Phys. Rev. E \textbf{59} (1999), R6263--R6266.

\bibitem{Sato95} M. Sato, M. Uwaha, \textit{Step Bunching as Formation of Soliton-like
Pulses in Benney Equation.} Europhys. Lett. \textbf{32} (1995), 639--644.

\bibitem{Bena93} I. Bena, C. Misbah, A. Valance, \textit{Nonlinear evolution of a terrace
edge during step-flow growth.} Phys. Rev. B \textbf{47} (1993), 7408--7419.

\bibitem{PierreLouis98b} O. Pierre-Louis, C. Misbah, \textit{Dynamics and fluctuations 
during MBE on vicinal surfaces. II. Nonlinear analysis.} Phys. Rev. B \textbf{58} (1998),
2276--2288.

\bibitem{PierreLouis98c} O. Pierre-Louis, C. Misbah, Y. Saito, J. Krug, P. Politi,
\textit{New Nonlinear Evolution Equation for Steps during Molecular Beam Epitaxy on 
Vicinal Surfaces.} Phys. Rev. Lett. \textbf{80} (1998), 4221--4224.

\bibitem{Kallunki00} J. Kallunki, J. Krug, \textit{Asymptotic step profiles from a 
nonlinear growth equation for vicinal surfaces.} Phys. Rev. E \textbf{62} (2000), 6229--6232.

\bibitem{Politi04} P. Politi, C. Misbah, \textit{When Does Coarsening Occur in the Dynamics
of One-Dimensional Fronts?} Phys. Rev. Lett. \textbf{92} (2004), 090601.

\bibitem{Danker03} G. Danker, O. Pierre-Louis, K. Kassner, C. Misbah,
\textit{Interrupted coarsening of anisotropic step meander.} Phys. Rev. E \textbf{68} (2003),
020601(R).

\bibitem{Kallunki04} J. Kallunki, J. Krug, \textit{Breakdown of step-flow growth in unstable
homoepitaxy.} Europhys. Lett. (2004), in press.




\end{thebibliography}
\end{document}